# Method for Extracting Patterns of Coordinated Network Attacks on Electric Power CPS based on Temporal-Topological Correlation

LEI WANG[1,2](Student Member, IEEE), ZHAOYANG QU[1,2](Member, IEEE), YANG LI[1](Senior Member, IEEE), KEWEI HU[3], JIAN SUN[3], KAI XUE[3] AND MINGSHI CUI[4]

[1]School of Electric Engineering, Northeast Electric Power University, Jilin 132012, China
[2]Jilin Engineering Technology Research Center of Intelligent Electric Power Big Data Processing, Jilin 132012, China
[3]State Grid Jilin Province Electric Power Supply Company, Changchun 130000, China
[4]EastInner Mongolia Electric Power Company, Hohhot 010000, China

Corresponding author: ZHAOYANG QU (e-mail: qzywww@neepu.edu.cn).

This work was supported by the Key Projects of the National Natural Science Foundation of China under Grant 51437003 and the Jilin science and technology development plan project of China under Grant 20180201092GX.

**ABSTRACT** In the analysis of coordinated network attacks on electric power cyber-physical system (CPS), it is difficult to restore the complete attack path, and the intent of the attack cannot be identified automatically. A method is therefore proposed for the extracting patterns of coordinated network attacks on electric power CPS based on temporal-topological correlation. First, the attack events are aggregated according to the alarm log of the cyber space, and a temporal-causal Bayesian network-based cyber attack recognition algorithm is proposed to parse out the cyber attack sequences of the same attacker. Then, according to the characteristic curves of different attack measurement data in physical space, a combination of physical attack event criteria algorithm is designed to distinguish the types of physical attack events. Finally, physical attack events and cyber attack sequences are matched via temporal-topological correlation, frequent patterns of attack sequences are extracted, and hidden multi-step attack patterns are found from scattered grid measurement data and information from alarm logs. The effectiveness and efficiency of the proposed method are verified by the testbed at Mississippi State University.

## I. INTRODUCTION

The "Ukrainian Blackout" in 2015, a landmark event in history in which a cyber attack was made on a power grid, fully confirms that cyber attack could cripple essential public systems. The SANS ICS information security organization has clearly stated that the cause of the incident was a coordinated network attack [1]. A coordinated attack can also look as though multiple attackers are working together to execute a distributed scan on many internal addresses or services. It is believed that probes of this nature have been developed in an attempt to elude the scan detection code present in many intrusion detection systems [2]. The destructiveness of coordinated network attacks is increasing with the widespread application of a large number of smart terminals and advanced measurement devices in smart grids [3].

In traditional information security technology, the intrusion process of the physical system is not considered, and it is difficult to effectively identify potential physical attack behavior [4]. Additionally, due to the combination of both cyber and physical attacks, the existing protection methods, such as intrusion detection system (IDS), firewalls, and abnormal data detection, lack effective correlation capabilities and cannot identify multi-step coordinated network attacks [5]. Therefore, there is an urgent need to study how to extract hidden multi-step attack patterns to reveal the complete process of intrusion behavior via the integration of physical grid operation information and cyber system alarm information.

Many scholars have carried out research on the pattern recognition and mining of coordinated network attack sequences. A mining algorithm for cascading failure model based on sequential pattern mining has been proposed by Trang *et al.* [6], in which the simulation analysis of a massive amount of grid operation data is conducted to effectively identify the system's cascading failure modes. A sequential pattern mining method that takes into account the degree of interest of sequential patterns and the utility value of each item in the sequence was proposed by Li Y *et al.* [7-8]. It presents an improvement over traditional sequential pattern mining, and uses sequence frequency as an important measure for patterns. A multi-stage coordinated attack analysis model based on the finite state machine and intelligent planning technology, which can identify the steps and paths of an attack, was established in the work by Gan w *et al.* [9-11]. The reference [12-13] showed how a previously published method based on Monte Carlo simulation could be enhanced to take into account time-dependent phenomena. The reference [14-16] studied transmission vulnerability based on the fault chain theory of security science, the cascading failure process and its generic features was described according to a fault chain. The reference [17] findings highlight the need to consider the load and coupling preference when designing robust interdependent networks.

However, the existing methods and models have some limitations. (1) In the analysis of the attack process, only the pattern analysis of the cyber network space or the physical grid space is performed; this diminishes the integrity of the cyber-physical attack process. (2) Supervised learning based on finite state machines and intelligent planning focuses on the attack state transition analysis of smaller-scale systems, which requires too much prior knowledge and complex rules. It is not practical for application in large-scale CPS systems. (3) The existing mining models of attack patterns do not comprehensively consider the impacts of physical grid constraints and cyber network attacks on electric power CPS systems, and there remains a lack of effective methods for identifying attack events.

Based on the existing research, the main purpose of this paper is to mine frequent attack patterns consisting of cyber attack sequences and physical attack events, and then to restore the attacker's attack process on the entire electric power CPS system. In summary, the contributions of this paper are as follows.

1) The fuzzy C-means and temporal causal Bayesian network are introduced to aggregate cyber system alarm information and extract cyber attack sequences, which significantly reduces the proportion of alarms. Multi-step attack sequences are obtained via credibility calculation, which improves the accuracy and the number of identified multi-step attack sequences.

2) A new physical attack event recognition method that combines physical criteria conditions with rule criteria conditions is proposed, and considers the variation law of the characteristic curves for power grid measurement data under different attacks. Additionally, three new key indicators are defined to improve the random forest (RF). Finally, the identification accuracy of physical attack events is improved.

3) For the first time, the temporal and topological relationship of cyber-physical components is introduced into frequent pattern mining, which combines cyber attack sequences and physical attack events. The frequent subsequence in the complete attack sequence can be extracted, which is the coordinated network attack pattern. This method effectively reduces the number of candidate sequences to be scanned, and no prior knowledge is required to set rules manually, which improves the efficiency of pattern mining.

The remainder of this paper is organized as follows. In Section II, related definitions of coordinated network attacks and the attack process problem formulation are given. The extraction method of coordinated network attack patterns is presented in Section III. The validation and performance of the proposed method are analyzed in Section IV. Finally, the paper is concluded in Section V.

## II. RELATED DEFINITIONS AND PROBLEM FORMULATION

### A. RELATED DEFINITIONS OF COORDINATED NETWORK ATTACKS

Coordinated network attacks on electric power CPS include two processes that occur in different spaces [18]. In the cyber system space, attacks will generate a large number of discrete alarm sequences; in the power system space, attack events will cause continuous changes in measurement data. Moreover, there is an attack conversion relationship between cyber attacks and physical grid attacks, and there are certain attack patterns of specific attack steps. Therefore, the relevant concepts of the process of coordinated network attacks are defined as follows.

**Definition 1.** Cyber attack sequence (*CAS*). *CAS* aim at attacking the cyber components in electric power CPS systems, which include routers, switches, computing devices, etc. The same attacker aims to obtain certain cyber system permissions, and implements a complete multi-step attack to trigger an alarm event sequence, which is defined as *CAS*. $CAS = [s_1^{CE_i} > s_2^{CE_2} > ... > s_m^{CE_m}]$, where $s_k^{CE_i}$ ($1 \leq k \leq m$) indicates an alarm event, which is represented by a 7-tuple [19], $s_k^{CE_i} = $ (*cid*, *time*, *src-ip*, *dst-ip*, *src-port*, *dst-port*, *sig_name*), the superscript *CEi* indicates the number of cyber components, and the subscript *k* is the type number of the cyber attack event, and the related event sequence is ordered in chronological order such that $s_i.time \leq s_j.time$, ($1 \leq i \leq j \leq m$).

**Definition 2.** Physical attack event (*PAE*). *PAE* aim at attacking physical components, such as relay protection devices, breaks, transmission lines, etc. Attacks that influence or damage the operation status of the power grid by tampering with measurement data or physical device configuration parameters are defined as $PAE = [e_1^{PE1}/e_2^{PE2}/...,/e_n^{PEn}]$, where the superscript $PEi$ indicates the number of physical components, and subscript $k$ is the type of physical attack event. $e_k^{PEn}$ $(1 < k < n)$ will cause abnormal changes in grid measurement (e.g., the voltage and current phase angle, amplitude, impedance, etc.). While $PAE$ satisfies $e_l \cap e_m = \emptyset$, $(0 < l, m < n)$, only single physical attack event behavior is considered in this paper, and multi-step physical attack events are not analyzed.

**Definition 3.** Attack Pattern ($AP$): A $CAS$ and $PAE$ that belong to the same combined attack sequence $[s_1^{CE1} > s_2^{CE2} ...> s_m^{CEm} > e_n^{PEn}]$ represent the complete attack path sequence. The attack process is closely related to the topological structure of the physical power grid and the information network. Therefore, under a specific electric power CPS topological structure, although the attack paths are different, there is an implicit correlation between these paths. The most frequently occurring subsequence in the attack path sequence are defined as $AP$, which is expressed as $[s_j^{CEj} > s_{j+1}^{CEj+1} > ... > s_{j+i}^{CEj+i} => e_k^{PEk}]$, $(1 \le j+i \le m, 1 \le k \le n)$.

### B. ANALYSIS OF COORDINATED NETWORK ATTACK PROCESS

The complete coordinated network attack process includes two phases, namely a cyber attack and physical attack, as illustrated in Fig. 1. During the cyber attack phase, cyber components are primarily attacked through network intrusion means to obtain certain control rights, such as via vulnerability scanning, brute-force attack, and network monitoring [20-22]. Based on this, during the physical attack phase, via the injection of false data or other means of tampering with the scheduling control instructions or attacking the physical system components, such attack behavior will cause the power system to fail or the line load to be reduced, thereby disrupting the normal operation of the power grid [23-25].

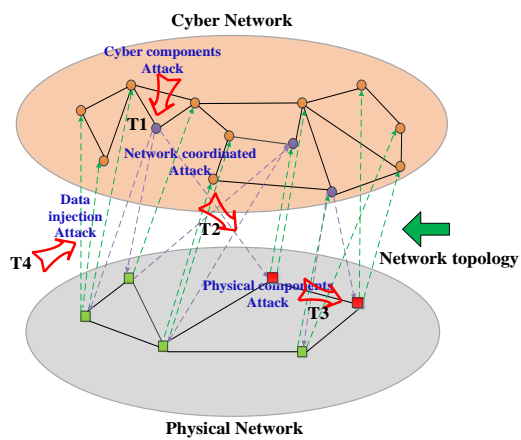

**FIGURE 1.** Coordinated network attack process.

Coordinated network attacks begin with the cyber attack, move through the security boundary of the cyber system and the physical system, and finally acts on the physical power grid [26]. It generally consists of multiple interrelated attack steps, the former step of which is often the condition for the latter step to occur [27].

Based on the above analysis of the main characteristics of coordinated network attacks, there are two kinds of relationships in the different attack stages.

1) Temporal relationship: If the premise of an attack on the physical grid is to obtain certain permissions, then the cyber attack that obtains this permission must happen first. In other words, to achieve a specific attack target, a cyber attack must be successfully carried out before the subsequent physical power grid attack, so the process determines the time range of the physical attack.

Suppose an attacker's attack path contains parts $CAS_i$ and $PAE_j$, which separately occur in the periods $[t_i, t_{i+n}]$ and $[t_j, t_{j+n}]$. If $CAS_i$ is a prerequisite for $PAE_j$, then $CAS_i$ must happen before $PAE_j$, that is, $[t_i, t_{i+n}] < [t_j, t_{j+n}]$.

2) Topological relationship: There is a topological relationship in the attack location. In a smart grid, cyber components ($CE$) and physical components ($PE$) are connected according to a certain topological structure. Therefore, the locations of attacked $CE$ determine the area that may be physically attacked. It is assumed that the physical system is an $n$-node power system, and the cyber system is an $m$-node communication and control network. The two networks represent the connection relationship through a ternary mapping table $O < CEs, PEs, R >$, that is $CEs$ represents a collection of $CE$ in a cyber-network, $PEs$ represents a collection of $PE$ in a physical-network, and $R$ represents the connection relationship between cyber-physical components. If $PE_i$ is connected to $CE$ ($CE_1, CE_2, ..., CE_n$), then there is a correlation between the cyber attack event sequence $[s_1^{CE1} > s_2^{CE2} > ... s_n^{CEn}]$ and the physical attack event $e_i^{PEi}$.

**Assumptions**

1) Because the dispatch data network uses a standard communication protocol, there is a possibility of being hacked.

2) After gaining the dispatcher's authority through attack methods such as permission elevation, the attacker can issue malicious load shedding instructions, leading to various types of grid accidents.

3) Cyber and physical attacks occur within a certain interval. When the interval exceeds a certain time limit, the attack components completed earlier may lose their influence on subsequent attacks due to human recovery or security early warning.

### III. EXTRACTION METHOD OF COORDINATED NETWORK ATTACK PATTERNS

#### A. OVERALL FRAMEWORK

The proposed framework for the extraction of coordinated network attack patterns is shown in Fig. 2. It consists of three main stages: the *CAS* recognition in the cyber system, the *PAE* recognition in the physical grid, and the *AP* extraction based on the temporal-topological correlation. The specific steps are as follows.

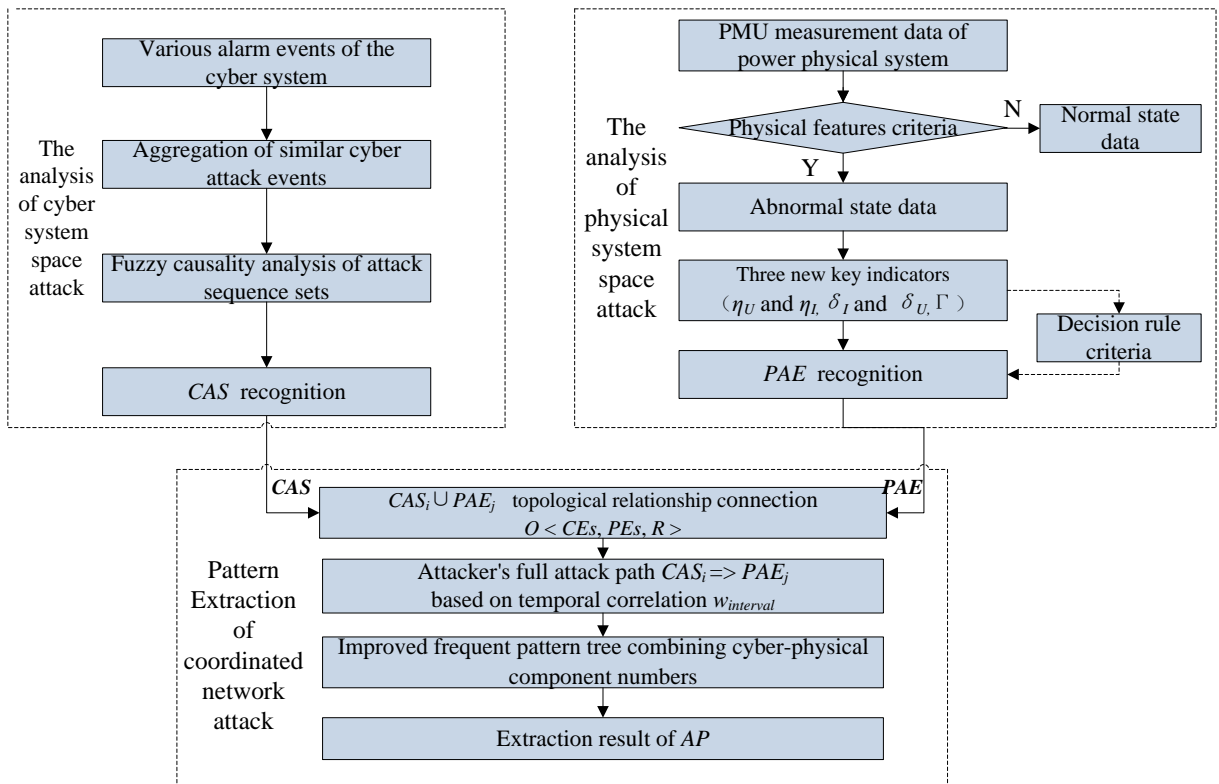

**FIGURE 2.** Framework of the attack pattern extraction method.

1) During the analysis phase of the attack process on the cyber system space, alarm events in the cyber system, such as the IDS, syslog, firewall log, and router log, is comprehensively used. According to the characteristics of cyber attack, the fuzzy c-means method is used to cluster the alarm events, therefore similar alarm events are aggregated together. Then, a temporal causal Bayesian network is used to perform a fuzzy causal analysis on the sequence set to construct a *CAS* set. Via credibility calculation, the *CAS* that belongs to the same attacker with the highest probability is recognized.

2) During the analysis phase of the attack process on the physical system space, the measurement data collected by the power management unit (PMU) is obtained. Physical feature criteria conditions are established based on the operating characteristic curves under different attacks of the physical power grid, and the normal state and abnormal state of the physical system are preliminarily judged. Then, three key indicators ($\delta_I$ and $\delta_U$ represent the phase difference change rate of voltage or current, $\delta_I$ and $\delta_U$ represent the amplitude fluctuation rate of voltage or current, $\tau$ represents the mutation coefficient of the negative/zero sequence) are defined to represent the law of attack behavior. The specific attack type decision rule criterion for the abnormal state dataset is established based on the improved RF algorithm, and the *PAE* is recognized by relying on the above rules.

3) During the extraction phase of the coordinated attack pattern, $CAS_i \cup PAE_j$ are connected based on the temporal-topological correlation. A dynamic sliding window $w_{interval}$ is then designed according to the temporal relationship to obtain the full attack path of the attacker. A coordinated network attack sequential pattern mining algorithm is proposed based on frequent pattern tree. According to the attack pattern threshold $\alpha$ and $\beta$, all coordinated network attack *APs* are extracted.

**B. CAS RECOGNITION BASED ON FUZZY FEATURE ANALYSIS**

1) CYBER ATTACK EVENT AGGREGATION

When an attacker launches an attack on a power information network, a large number of warning traces will be left in the cyber components, including the IDS, router, and firewall. Considering that the alarm events $s_1, s_2, ..., s_k$ of multi-step cyber attack activities have a certain internal relationship in time and space, it can restore the steps and purpose of the attacker. However, these alarm logs are scattered throughout different cyber components, most of them are low-level isolated events that lack relevance and have high false positive and false negative rates; therefore, traditional attack methods cannot be used to directly divide attack sequences.

For this reason, the idea of fuzzy cluster analysis is adopted, and a fuzzy C-means clustering algorithm that takes into account the characteristics of cyber attacks based on the characteristics of the attacked cyber components is proposed.

Let $u_{ik}$ represent the attack activity $x_i$ belong to the membership degree of alarm event $s_k$; $0 \leq u_{ik} \leq 1$, and $\sum_{i=1}^{n} u_{ik} = 1$. The objective function is then defined as follows [28].

$$J(U,V) = \sum_{k=1}^{n} \sum_{i=1}^{c} u_{ik}^m d_{ik}^2 \quad (1)$$

where $U = (u_{ik})_{i \times n}$ is the membership degree matrix, $m$ is the number of clusters, $d_{ik} = \|x_k - v_i\|$, and $J(U,V)$ is the sum of the weighted squared distance from the attack activity to the cluster center in each alarm event. Additionally, the weight of attack activity $x_k$ belongs to the $m$-th power of the membership degree of the type $i$-th attack event. The clustering criterion of the algorithm is to calculate $U$ and $V$ to make $J(U, V)$ have the minimum value, and indicates that related attacks $x_k$ have similar attack characteristics and belong to the same type of attack.

### 2) RECOGNITION OF CYBER ATTACK SEQUENCE
Based on the aggregate calculation of network attack alarm events, two factors must be considered to identify the *CAS* belonging to the same attacker. First, due to noise interference and the failure of the attacked component itself, false alarms and the phenomenon of underreporting are difficult to avoid. Second, the causal relationship between the attack sequence and the alarm event usually has a certain timeliness and time uncertainty. The *CAS* should be formed within a certain period of time after different alarm events, but it is difficult to accurately determine a clear time interval.

This paper proposes a method based on the temporal causal Bayesian network to identify *CASs* of the same cyber attack sequence. As presented in Fig. 3, the *CAS* recognition method is a two-layer directed acyclic graph, the root node of which represents the alarm event, and the leaf node of which represents the cyber attack sequence. It can be described by a quadruple $G=\{S,CAS',E,K\}$, where $S$ represents an alarm event node set, $CAS'$ represents a fuzzy subset of cyber attack sequences, $E$ is the edge set that the cause event node points to the result event node, $E_{ij} \in E$ indicates that alarm events $S_i$ and attack result events $CAS_j$ have a causal relationship, and $K$ represents the set of states that have causality. Any $K_{ij} \in K$ can be the state of the side switch.

The specific steps of the method are as follows.

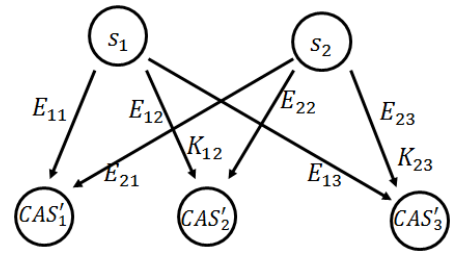

**FIGURE 3.** Model of temporal causal Bayesian network cyber attack sequence.

**Step1:** When multiple alarm events $s_1, s_2,...,s_k$ can form the same *CAS*, i.e., when the attack sequence contains multiple root nodes, each root node event can independently trigger the cyber attack.

**Step2:** After performing fuzzy operations on all the temporal-causal relationships of the *CAS*, a fuzzy subset of the cyber attack sequence $CAS' = [s_1^{CE1} > s_2^{CE2} > ... > s_m^{CEm}]$ can be obtained, in which different permutations of alarm events constitute different fuzzy subsets *CAS'*.

**Step3:** The fuzzy subset of the *CAS* contains all possible combinations of the alarm events forming the attack sequence. As shown in the example in Fig. 3, there are three kinds of *CASs* that may attack the sequence set: $CAS' = \{CAS'_1, CAS'_2, CAS'_3\}$, where $CAS'_1 = (s_1^-, s_2^+)$, $CAS'_2 = (s_1^+, s_2^-)$, and $CAS'_3 = (s_1^-, s_2^-)$.

**Step4:** Probability calculations are used to determine which attack sequence $CAS'_i$ has the greatest credibility. For any attack sequence $CAS'_i \in CAS'$, its credibility is calculated as follows.

$$Bel(CAS'_j) = \max\left\{\alpha \prod_j \delta(CAS|S,K) \prod_{i:s_i=s_i^+} \pi_i \prod_{i:s_i=s_i^-} (1-\pi_i)\right\} \quad (2)$$

Where $\alpha$ is the normalization constant, and $\pi_i$ is the $s_i$ prior probability of occurrence. The maximum value of $Bel(CAS'_j)$ indicates the alarm event combination $[s_1^{CEi1}, s_2^{CEi2},...,s_m^{CEim}]$, which is an accurate sequence of actual $CAS_j$.

## C. PAE RECOGNITION BASED ON COMBINED CRITERIA
### 1) PHYSICAL FEATURE CRITERIA CONDITIONS
The measurement data of the physical power grid contains many attributes, including the voltage, current, phase angle, positive sequence, negative sequence, and zero sequence. Via analysis, it is found that under different types of attacks, there are obvious differences in the characteristic curves of these measurement data. The features can be summarized as follows.

*Attack type 1:* Data injection attack. The measured value is maliciously modified to disguise a normal fault, causing the operator to mistakenly believe that a short-circuit fault occurred and to issue a removal instruction. Fig. 4 presents a recording of the time interval from 42 to 92, in which malicious data tampering was performed, resulting in the phase differences between *A-C* three-phase voltages not being 120 degrees.

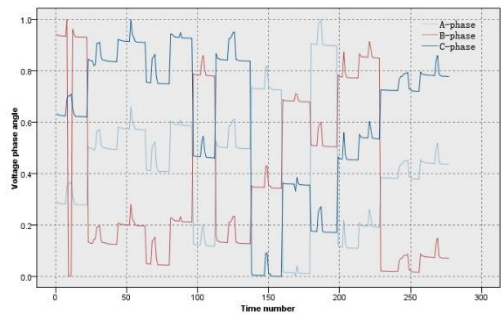

**FIGURE 4.** Characteristic curve of voltage phase angle.

*Attack type 2:* Command injection attack of remote switch. An attacker sends a malicious operation command to the relay, causing the circuit breaker position to change. Fig. 5 presents the positive-negative-zero sequence current amplitude changes in the relay after the attacker injected a malicious disconnect command in the time intervals from 36 to 38 and from 153 to 158.

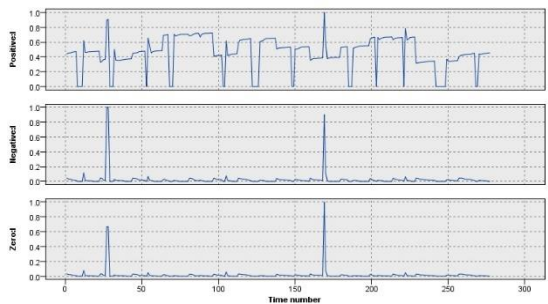

**FIGURE 5.** Characteristic curve of positive-negative-zero sequence.

*Attack type 3:* Relay setting parameter tampering attack. The impedance relay main function is to measure the distance from the short-circuit point to the protection installation. In the attack state, because the protection distance parameter is tampered with, the protection has no effect. Fig. 6 displays the attacker tampering with relay protection parameters in the time intervals from 25 to 60, and presents the change in impedance angle *ZH* between the attack failure and the normal failure.

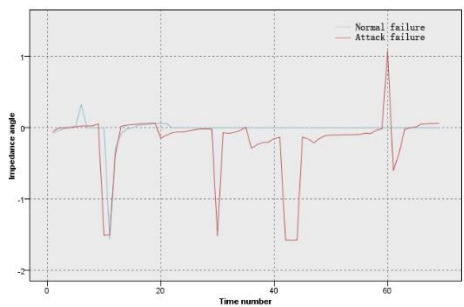

**FIGURE 6.** Characteristic curve of the impedance phase angle.

Based on the analyses of the characteristic curves of these three types of attack states, the physical feature criteria for establishing abnormal conditions according to the operating characteristics of the physical power grid are listed in Table 1.

**TABLE 1.** Physical feature criteria for abnormal conditions.

| Abnormal number | Physical feature criteria |
|---|---|
| Abnormal feature 1 | $\forall t \in [t_0, t_0 + \triangle T]$, $\|(\varphi_A(t) - \varphi_B(t)) - (\varphi_B(t) - \varphi_C(t))\| \geq \varepsilon_1$ or $\|(U_A(t) - U_B(t)) - (U_B(t) - U_C(t))\| \geq \varepsilon_2$ or $\|(I_A(t) - I_B(t)) - (I_B(t) - I_C(t))\| \geq \varepsilon_3$ |
| Abnormal feature 2 | $\forall t \in [t_0, t_0 + \triangle T]$, $I_-(t) \neq 0$ or $I_0(t) \neq 0$ |
| Abnormal feature 3 | $\forall t \in [t_0, t_0 + \triangle T]$, $\triangle z = \left\|\frac{z(t_1) - z(t_0)}{z(t_1)}\right\| > \varepsilon_4$ |

In Table 1, $\varphi(t)$, $U(t)$, $I(t)$, and $z(t)$ respectively represent the phase angle, voltage amplitude, current amplitude, and impedance amplitude measured at time *t*, $I_-(t)$ and $I_0(t)$ respectively represent negative and zero sequence currents, $\Delta z$ represents the rate of change of impedance, and $\varepsilon_1$, $\varepsilon_2$, $\varepsilon_3$, and $\varepsilon_4$ denote the allowable error ranges [29].

2) RECOGNITION OF PHYSICAL ATTACK EVENTS

To accurately identify the specific attack type of the abnormal state, three key indicators are defined according to the change laws of the measurement characteristics under the attack.

*Indicator 1:* Phase difference change rate of voltage or current. Taking a fixed interval window *n*, to reflect the average deviation of the three-phase current or voltage.

$$\eta_U = \frac{\sum_{t=1}^{n}(|\varphi_u^A(t) - \varphi_u^B(t)| - |\varphi_u^B(t) - \varphi_u^C(t)|)}{n} \quad (3)$$

$$\eta_I = \frac{\sum_{t=1}^{n}(|\varphi_i^A(t) - \varphi_i^B(t)| - |\varphi_i^B(t) - \varphi_i^C(t)|)}{n} \quad (4)$$

*Indicator 2:* The amplitude fluctuation rate of voltage or current reflects the fluctuation degree of the three-phase current or voltage at a certain moment.

$$\delta_I = \frac{\max(I_{A \vee B \vee C} - \frac{I_A + I_B + I_C}{3})}{\frac{I_A + I_B + I_C}{3}} \times 100\% \quad (5)$$

$$\delta_U = \frac{\max(U_{A \vee B \vee C} - \frac{U_A + U_B + U_C}{3})}{\frac{U_A + U_B + U_C}{3}} \times 100\% \quad (6)$$

*Indicator 3:* The mutation coefficient of the negative/zero sequence. When the system fails or attacks, it usually breaks down the current negative and zero sequence amplitude. Therefore, detecting these two variables that should not occur in the normal state can be used as a basis for judging whether the system is abnormal.

Setting the window length $\Delta t$, If $I_0(t)$ or $I_-(t)$ have a change in the adjacent $\Delta t$ interval is more than 20% [30], it may be considered as an abnormal value.

$$\tau(t) = \begin{cases} 1, & |I_{0or\_}(t) - I_{0or\_}(t-\Delta t)| > 0.2 \times I_{0or\_}(t-\Delta t) \text{ and} \\ & |I_{0or\_}(t+\Delta t) - I_{0or\_}(t)| > 0.2 \times I_{0or\_}(t) \\ 0, & |I_{0or\_}(t+\Delta t) - I_{0or\_}(t)| < 0.2 \times I_{0or\_}(t) \end{cases} \quad (7)$$

The above three indicators are merged, a physical attack event recognition algorithm based on RF, called *PAR-RF*, is proposed to realize the classification detection of abnormal measurement data. The main steps of the algorithm are as follows.

**Step1:** The original physical grid measurement dataset $W$ is obtained according to the established physical feature criteria condition table to perform a preliminary determination of the normal state and the attack state. Additionally, a candidate attack dataset that may have attack events $W'$ is generated.

**Step2:** Supposing that $W'$ contains $m$ samples and each sample has $X$ features, a random forest containing $T$ decision trees must be trained, Bootstrap is used for sampling, and $T$ training sets $K$ of size $n$ ($n < m$) are obtained from $W'$.

**Step3:** In each training set $K$, the phase change rates of voltage and current $\eta_U$ and $\eta_I$, the amplitude fluctuation rates of voltage and current $\delta_I$ and $\delta_U$, and the mutation coefficient of the negative/zero sequence $\tau$ are calculated. Additionally, the feature extraction method is used to select important features $x'$ according to the feature importance value ($x' < X$), and combined with the three key indicators, namely ($x'$, $\eta_U$, $\eta_I$, $\delta_I$, $\delta_U$, $\tau$).

**Step4:** CART is used to build a classification decision tree, and steps 2 and 3 are repeated to build $T$ decision trees, i.e., $T$ classifiers. Rules are then generated to determine the type of attack. Finally, the voting results of the $T$ classifiers are used to determine the final attack *PAE* event category $e_i$.

### D. AP EXTRACTION BASED ON TEMPORAL-TOPOLOGICAL CORRELATION

Based on the captured *CAS* and *PAE*, the temporal and topological relationships between the physical and cyber components are used as correlated conditions. The frequent pattern tree (FP-Tree) is improved via the time scale and topology, and a coordinated network attack sequential pattern mining algorithm (NSPMA) is proposed. By establishing a topology frequent pattern tree (TFP-Tree), the coordinated *AP* of the electric power CPS network is mined. The steps of the NSPMA method are as follows.

**Step1:** Cyber-physical attack sequential connection based on topology. According to the topological relationship $O < CEs$, $PEs$, $R >$, the *CAS* set $\{CAS_1, CAS_2,...,CAS_m\}$ and the *PAE* set $\{PAE_1, PAE_2,...,PAE_n\}$ are traversed, and if there is a connection relationship, operation $CAS_i \cup PAE_j$ is performed.

**Step2:** The complete attack sequence for each attacker is obtained based on a new dynamic time window. The cyber-physical attack sequence is sorted according to the attacker's *AID* as the primary key and the *timestamp* as the secondary key. To avoid dividing the sequence of the same attacker into different sequences, the dynamic time window $w_{interval}$ in the attack scene is designed. The attack sequence is then segmented to obtain the complete attack sequence database (*AD*) for all attackers.

For $CAS_i$ (which occurs at $[t_i, t_{i+n}]$) and $PAE_j$ (which occurs at $[t_j, t_{j+n}]$), the attack interval is $\Delta t = |t_j - t_{i+n}|$. In the relevant literature [31], when $\Delta t > 240$ seconds, after the attack step interval exceeds 240 seconds, the attack result will be invalid, even without any threat. The maximum upper limit of $w_{interval}$ is set to 240 seconds. When $\Delta t < 240$ seconds, $w_{interval}$ is dynamically calculated and determined according to the average time interval of each type of attack sequence.

$$w_{interval} = \begin{cases} 240, & \Delta t > 240 \text{ seconds} \\ \dfrac{\sum_{i=1}^{l} n_i * t_k}{\sum_{i=1}^{l} n_i}, & \Delta t < 240 \text{ seconds} \end{cases} \quad (8)$$

where $t_k$ represents a time interval in which a certain type of attack sequence $CAS_i => PAE_j$ may occur among different attackers, and $n_i$ represents a cumulative number of occurrences of the $CAS_i => PAE_j$ sequence in each $t_k$ interval.

**Step3:** Construct attack sequence TFP-Tree. By setting the threshold $\alpha$ for attack pattern support, the *AD* is scanned to count each of the attack items, and the *CE* or *PE* component number is recorded simultaneously. Only the frequent attack items greater than $\alpha$ are retained, and the support items are sorted in descending order. Then, by scanning the attack sequence of each attacker in the *AD*, the sorted frequent items in each attack sequence are obtained, and the attack sequence TFP-Tree is established, as shown in Fig. 7.

$$\alpha = \frac{\text{Number of attackers on } [S_1 > S_2...S_{n-1} => e_i]}{\text{Total number of attackers}} \quad (9)$$

**Step4:** The TFP-Tree is mined to find extremely frequent sequences and generate *APs*. Each item in the header of table is traversed in turn, all existing attack paths are extracted according to the node list, and the subschema base is calculated to build a conditional FP tree. Finally, all *APs* are obtained according to the set attack sequence mode confidence $\beta$ threshold.

$$\beta = \frac{\text{Number of attacks on [antecedent } S_1 > S_2...S_{n-1} \cap \text{consequent } e_i]}{\text{Number of antecedent attacks } [S_1 > S_2...S_{n-1}]}$$
$$(10)$$

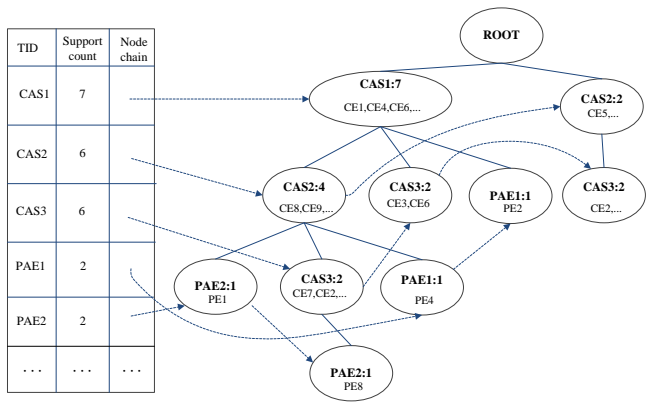

**FIGURE 7.** TFP-Tree example.

## IV. EXPERIMENT AND ANALYSIS

The testbed of Mississippi State University was utilized to analyze and verify the proposed method of coordinated network attack pattern extraction [32]. The experimental connection topology is illustrated in Fig. 8. The physical grid system contains the following components: 2 power generators *G1* and *G2*, 4 intelligent electronic devices (IED) *R1* through *R4*, they can switch the breakers on or off. 4 circuit breakers *BR1* to *BR4*. *R1* controls *BR1*, *R2* controls *BR2* and son on accordingly. 2 grid lines *L1* and *L2*; the components are numbered from *PE1* to *PE12*. The cyber system contains the following components: a router, switch, snort, syslog server, open PDC, computer terminal, software system, etc. The components are numbered from *CE1* to *CE9*.

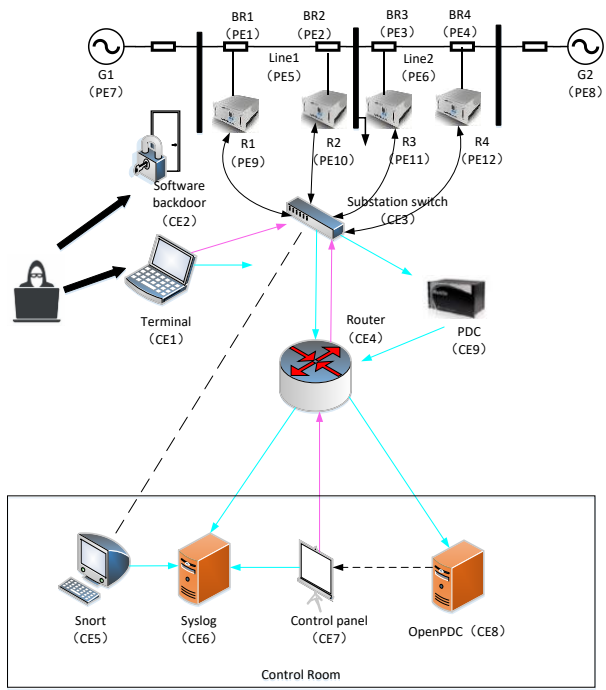

**FIGURE 8.** Testbed topology.

### A. VALIDATION AND ANALYSIS OF CAS RECOGNITION

A total of 28,901 alarm data was generated in the deployed snort, syslog servers, routers, and switches. A fuzzy C-means clustering algorithm that takes into account cyber attack characteristics was verified to aggregate network attack events. In the experiment, the attackers were divided by *IP address*. Membership degree was calculated based on the continuous variables such as *timestamp*, and *sig_name*, and the alarm events to which attack activities belong were determined by the values. There were 12,360 alarm events after calculation, and the results are presented in Fig. 9, which displays the changes in the numbers of the top 8 cyber alarm events before and after aggregation.

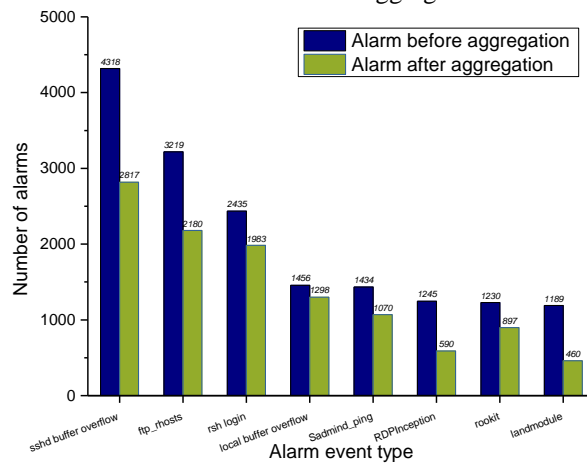

**FIGURE 9.** Comparison of the numbers of original alarms and aggregated alarms.

The events were numbered from $s_1$ to $s_8$, and the specific event information is presented in Table 2.

**TABLE 2.** Meanings of some cyber alarm events.

| Serial number | Attack type | Description |
|---|---|---|
| $s_1$ | sshd buffer overflow | Remotely obtain root permissions on the target host. |
| $s_2$ | ftp_rhosts | Gain the trust of the target host. |
| $s_3$ | rsh login | Get user permissions on the target host. |
| $s_4$ | local buffer overflow | Get root privileges on localhost. |
| $s_5$ | sadmind_ping | Search for active hosts to target attacks. |
| $s_6$ | RDP Inception | Wait for the intranet administrator to initiate an RDP connection request and repeatedly obtain the management control shell of all devices. |
| $s_7$ | Rookit | Install on the target component and hide to realize the functions of hiding and manipulating and collecting data. |
| $s_8$ | Landmodule | Send spoofed data packets to disable target devices without corresponding protection mechanisms. |

Based on this clustering data, a temporal causal Bayesian network model was used to identify the attack sequences belonging to the same attacker, and a total of 420 cyber attack sequences was obtained. Based on the maximum confidence probability calculation, a total of 138 actual attack activity sequences was identified. Due to space limitations, only the top 10 cyber attack sequences with high credibility are listed, as presented in Table 3.

**TABLE 3.** Top 10 cyber attack sequences with high credibility.

| Serial number | Cyber attack sequence set | Reliability |
|---|---|---|
| $CAS_1$ | $s_5^{CE1} > s_6^{CE3} > s_1^{CE2}$ | 98.4% |
| $CAS_2$ | $s_7^{CE1} > s_5^{CE3} > s_4^{CE2}$ | 97.6% |
| $CAS_3$ | $s_7^{CE7} > s_2^{CE4} > s_4^{CE3}$ | 85.4% |
| $CAS_4$ | $s_6^{CE5} > s_3^{CE4} > s_4^{CE3}$ | 94.3% |
| $CAS_5$ | $s_5^{CE7} > s_7^{CE4} > s_3^{CE3}$ | 91.8% |
| $CAS_6$ | $s_8^{CE8} > s_7^{CE7} > s_2^{CE4} > s_4^{CE3}$ | 82.5% |
| $CAS_7$ | $s_6^{CE1} > s_8^{CE6} > s_2^{CE4} > s_4$ | 81.7% |
| $CAS_8$ | $s_7^{CE5} > s_5^{CE4} > s_8^{CE3} > s_4^{CE2}$ | 80.6% |
| $CAS_9$ | $s_6^{CE1} > s_5^{CE4} > s_8^{CE6} > s_2^{CE5} > s_4^{CE3}$ | 77.1% |
| $CAS_{10}$ | $s_8^{CE6} > s_5^{CE7} > s_6^{CE4} > s_2^{CE3} > s_4^{CE2}$ | 76.6% |

As is evident from Table 3, the maximum credibility of a 3-step attack was 98.4%, the maximum credibility of a 4-step attack was 82.5%, and the maximum credibility of a 5-step attack was 77.1%. As the length of the identified cyber attack sequence increased, the credibility decreased significantly. This is because when the length of the attack sequence increases, the attacker's intention to attack is uncertain. Various tentative attacks will affect the accuracy of the identification rate.

The cyber attack sequence recognition method proposed in this paper and other methods proposed in previous research [33] are aimed at the mining of network alarm logs of multi-step cyber attack sequences. Thus, to further verify the effectiveness of the method proposed in this paper, an experimental comparison with related methods was carried out. The aggregated 12,360 alarm event data were used in the experiment, and were divided into 10 groups of test data sets. The experimental results are presented in Fig. 10.

The experimental results demonstrate that the proposed method exhibited obvious advantages in both the number of recognized attack sequences and the accuracy of recognition. This is because the proposed method establishes an optimization objective function to determine the membership of each sample point to all class centers, thereby effectively reducing the number of redundant types of alarm information and making up for the shortcomings of the other algorithms, which only cluster by the time attribute or attack type. It is therefore more in line with the actual situation of multi-step attack implementation. On the contrary, the intelligent planning method pre-defines the prerequisites and consequences of the attack steps, and the alarm correlation method uses the time window to segment the alarm sequence in sections, the randomness and uncertainty of the attack sequence events are ignored.

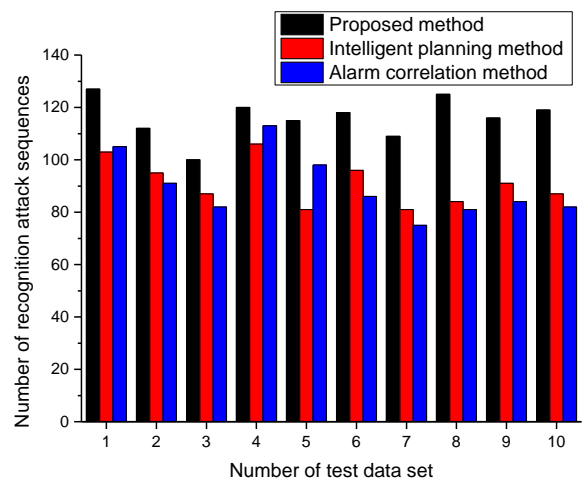

(a) Number of recognized attack sequences

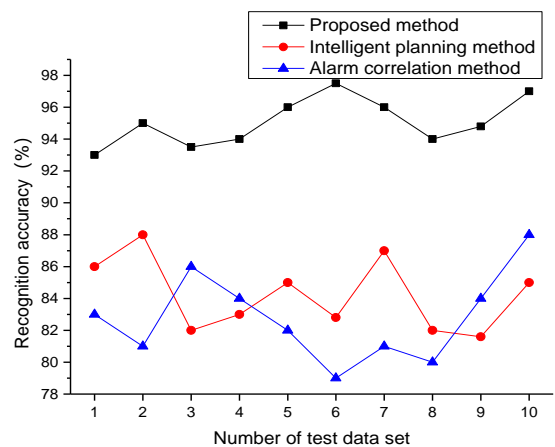

(b) Recognition accuracy

**FIGURE 10.** Performance comparison of the proposed method, intelligent planning method, and alarm correlation method.

### B. VALIDATION AND ANALYSIS OF PAE RECOGNITION

The physical grid data set was collected by 4 PMUs, and includes 28 types of attack scenarios and 9 types of normal, short-circuit, and maintenance scenarios. The numbers and event descriptions of all 37 scenarios are listed in Table 4.

**TABLE 4.** Physical attack event descriptions.

| Event number | Description |
|---|---|
| $e_{41}$ | Normal |
| $e_1$-$e_6$ | short-circuit fault from 10-19%, 20-79%, 80-90% on L1 or L2 |
| $e_{13}$, $e_{14}$ | Line1 or Line2 maintenance |
| $e_{15}$-$e_{20}$ | remote command injection attack on R1 to R4 |
| $e_{21}$-$e_{30}$, $e_{35}$-$e_{40}$ | relay setting change attack from 10-19%, 20-90%, 10-49%, 50-79% on R1 to R4 |
| $e_7$-$e_{12}$ | measuring data injection attacks from 10-19%, 20-79%, 80-90% on L1 or L2 |

The data set contains 128 features with a total of about 70,000 pieces of data, and is divided into 15 sub-data sets. The features are explained in the Table 5. There are 29 types of measurements from each PMU, so the 4 PMUs have 116 measurement columns. There are 12 columns for control panel logs, snort alerts and relay logs. The last column is the marker.

**TABLE 5.** Data set features descriptions.

| Feature | Description |
|---|---|
| PA1:VH – PA3:VH | Phase A - C Voltage Phase Angle |
| PM1: V – PM3: V | Phase A - C Voltage Phase Magnitude |
| PA4:IH – PA6:IH | Phase A - C Current Phase Angle |
| PM4: I – PM6: I | Phase A - C Current Phase Magnitude |
| PA7:VH – PA9:VH | Pos. – Neg. – Zero Voltage Phase Angle |
| PM7: V – PM9: V | Pos. – Neg. – Zero Voltage Phase Magnitude |
| PA10:VH - PA12:VH | Pos. – Neg. – Zero Current Phase Angle |
| PM10: V - PM12: V | Pos. – Neg. – Zero Current Phase Magnitude |
| F | Frequency for relays |
| DF | Frequency Delta (dF/dt) for relays |
| PA:Z | Appearance Impedance for relays |
| PA:ZH | Appearance Impedance Angle for relays |
| S | Status Flag for relays |

The physical attack event recognition method was verified, and an initial random forest with a size of 100 was then constructed. The decision rules were generated from the training data set, and the top 5 rules for accuracy are listed in the Table 6. The index of each rules is in the form of "R#-Signal Reference" that indicates a type of measurement from a PMU specified by "R#". For example, R3-PM7:V means Pos voltage phase magnitude measured by PMU R3.

**TABLE 6.** Top 5 decision rules for accuracy.

| Decision rule | Most frequent category | Rule accuracy |
|---|---|---|
| (R3-PM7:V <= 130130.2713) and (R2-PM6:I <= 383.337036) and (R2-PA3:VH > 48.856111) and (R2-PM5:I <= 486.88949) | 24 | 0.978 |
| (R4-PM6:I <= 377.38971) and (R1-PM5:I > 357.98005) and (R1-PA6:IH > 49.841599) and (R2-PM5:I <= 510.32757) | 24 | 0.962 |
| (R4-PM2:V > 131127.0625) and (R2-PM3:V <= 130431.1505) and (R2-PM4:I <= 485.115051) and (R4-PA3:VH <= -13.81387799) | 12 | 0.957 |
| (R2-PA2:VH <= -156.864385) and (R4-PM5:I > 380.8688) and (R3-PA2:VH <= -34.61811) and (R3-PA5:IH > -125.02512) | 36 | 0.951 |
| (R3-PM5:I <= 424.99831) and (R2-PA2:VH > -156.864385) and (R2-PM5:I > 362.709045) and (R2-PA3:VH > 48.8561) | 13 | 0.948 |

According to the classification rules, the accuracy analysis experiment of physical attack type recognition was performed. The experiment was performed using PAR-RF and the original RF on the test data set. The size of the initial random forest scale was 20. To ensure the stability of the experimental results, the experiment was repeated 20 times on each data set. The recognition capabilities of the two methods are presented in Fig. 11.

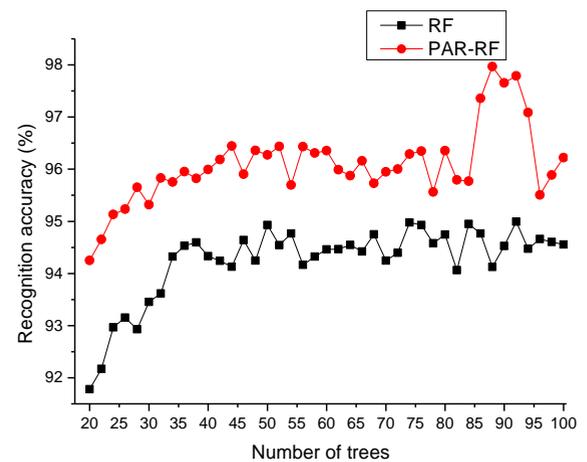

**FIGURE 11.** Comparison of classification accuracy under forests of different scales.

The experimental results reveal that PAR-RF exhibited better overall classification accuracy than RF due to the addition of the key indicators for classification feature selection. The best classification accuracy was 97.9%, and the forest size was 88.

Additionally, a test data set was randomly selected from 15 sample data sets. It contained a total of 5,069 samples, of which 4887 test classifications were correct, accounting for 96.41%, and 182 were misclassified, accounting for 3.59%. The classification of 37 scenarios was analyzed via the confusion matrix performance, as presented in Fig. 12. The main diagonals of the matrix were classified correctly, the others were misclassified.

The accuracy and false positive rates of the 37 attack scenarios were calculated according to the confusion matrix, as presented in Fig. 13.

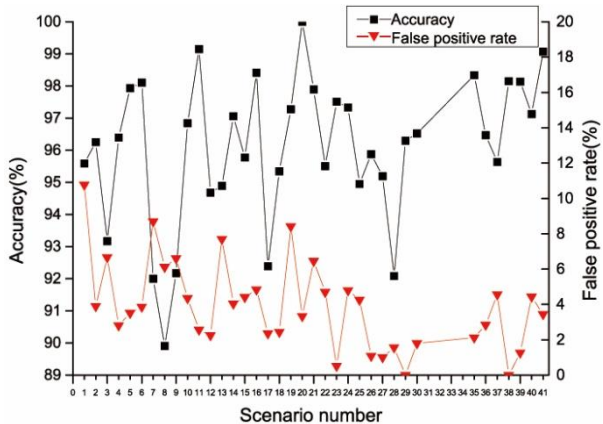

FIGURE 13. Accuracy and false detection rate of 37 scene recognitions.

accuracy rates were $e_{20}$, $e_{11}$, and $e_{16}$; the respective recognition accuracy rates were 100%, 99.06%, and 99.03%, and the respective false detection rates were 0.03%, 0.02%, and 0.04%. The top three misdetection rates for attack scenarios were $e_1$, $e_7$, and $e_{21}$; the respective misdetection rates were 0.11%, 0.09%, and 0.08%, and the respective accuracy rates were 95.6%, 92%, and 97.8%.

The causes of error are analyzed. $e_1$ is a normal short-circuit fault of Line1 at 10%-19% of the position, and the attack events $e_7$ and $e_{21}$ were tampering via malicious data injection or relay R1 parameter settings, resulting in a positive sequence of Line1 short-circuit fault at 10%-19%. Additionally, the negative/zero sequence amplitude and phase angle change were similar, which could easily cause confusion in recognition.

The experimental results demonstrate that the proposed method has a better recognition effect on single or combined relay remote command injection attacks and data injection attacks. The top three attack scene recognition

FIGURE 12. Confusion matrix for the recognition of 37 scenarios.

### C. VALIDATION AND ANALYSIS OF ATTACK PATTERN EXTRACTION

*CASs* and *PAEs* were recognized based on the reported experiments. The NSPMA algorithm was used to mine frequent item sets by establishing a TFP-Tree, $\alpha=30$，$\beta=30$；from which the sequential pattern of the coordinated network attack was extracted.

Via experimental verification, a total of 461 complete attack paths were obtained, and 22 frequent attack patterns were mined. The results are sorted in descending order according to the number of sequences that occur in each pattern, as shown in the Fig .14. This proved that frequent attack patterns account for only a small percentage.

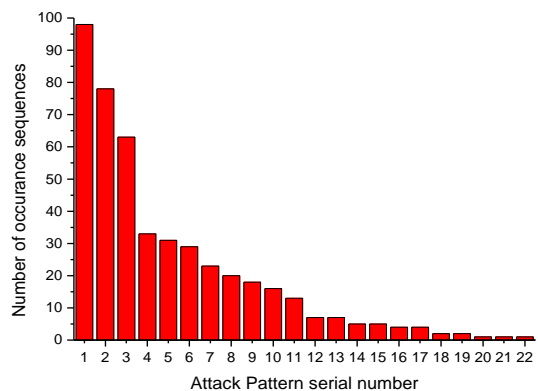

FIGURE14. Descending statistics of sequence number in attack pattern

A total of 6 frequent coordinated network attack sequential patterns with a confidence level of greater than 90% were found, as shown in Table 7. For example, the pattern $[s_5^{CE4} > s_1^{CE3} => e_{21}^{PE2}]$ indicates that when the cyber component *CE4* is attacked by a Sadmind Ping ($s_5$), and then *CE3* is attacked by an sshd buffer overflow ($s_1$), the physical component *PE2* suffers a relay setting change attack ($e_{21}$) with a confidence of 93.1%, support of 97.5%.

TABLE 7. Coordinated network attack pattern with top 6 confidence.

| Serial number | Coordinated network attack pattern | Confidence | Support |
|---|---|---|---|
| 1 | $[s_5^{CE1} > s_6^{CE3} > s_3^{CE2} => e_{16}^{PE2}]$ | 94.1% | 98.6% |
| 2 | $[s_5^{CE4} > s_1^{CE3} => e_{21}^{PE2}]$ | 93.1% | 97.5% |
| 3 | $[s_7^{CE7} > s_5^{CE4} > s_3^{CE3} => e_7^{PE11}]$ | 92.6% | 93.5% |
| 4 | $[s_6^{CE6} > s_8^{CE4} > s_4^{CE3} => e_{17}^{PE1}]$ | 91.3% | 92.7% |
| 5 | $[s_6^{CE4} > s_2^{CE3} => e_{27}^{PE9}]$ | 91.6% | 94.6% |
| 6 | $[s_3^{CE4} > s_4^{CE3} => e_9^{PE3}]$ | 90.5% | 95.4% |

By summarizing the results of these attack patterns, three typical network cooperative attack patterns can be determined.

1) The data injection attack patterns mainly focus on network monitoring, packet read and write (such as the Netcat tool) attack events, and the injection of maliciously tampered-with data to affect the dispatcher's decision, such as the following pattern: $[s_7^{CE7} > s_5^{CE4} > s_3^{CE3} => e_7^{PE11}]$.

2) The command injection attack pattern mainly uses password brute force cracking and privilege elevation attacks. After gaining the root privileges of the dispatcher, malicious load shedding attacks on the physical grid are performed, such as the following pattern: $[s_6^{CE6} > s_8^{CE4} > s_4^{CE3} => e_{17}^{PE1}]$.

3) The attack pattern of tampering with relay setting mainly targets router or switch security vulnerabilities. The relay device is invaded through a network connection to maliciously modify the protection distance parameter, which causes the relay to refuse to operate when a fault occurs, such as the following pattern: $[s_6^{CE4} > s_2^{CE3} => e_{27}^{PE9}]$.

To fully verify the efficiency of the NSPMA algorithm proposed in this paper, four representative sequential pattern mining algorithms, namely Prefix Span, Free Span, GSP, and Spade, were selected for comparison experiments.

Under the same conditions, according to the different support levels set, each algorithm was independently run 20 times, and the average running time was taken as the calculation result. The time efficiency comparison of the four algorithms is presented in Fig. 15.

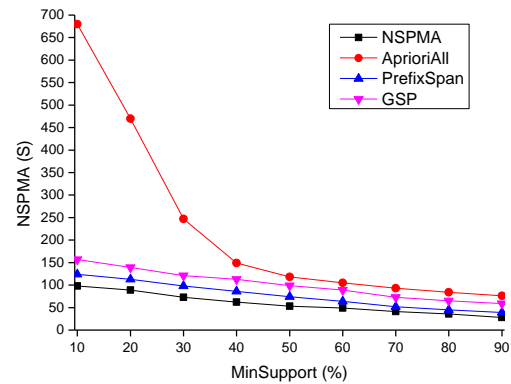

FIGURE 15. Algorithm running times with different support thresholds.

The results demonstrate that with the increase of the support threshold, the running time of each algorithm will decrease rapidly. The reason for this is that as the support threshold increases, the number of frequent pattern sequences that meet the threshold will decrease, and the running time will also decrease. In terms of running time, NSPMA was found to be superior to the other algorithms. This is because the connection topology of components is considered in the frequent item set comparison process, i.e., item sets that have no relationship will be directly pruned without the need for connection operations; this avoid a large number of useless candidate sequence item sets, thereby effectively improving the efficiency of pattern mining.

## VI. CONCLUSION AND FUTURE WORK

This paper proposes a new method for the automatic mining of attack patterns from measurement data and information alarm logs based on the characteristics of coordinated network attacks that occur in physical space and cyber space, and the temporal and topological correlation between each attack step. The proposed method can restore the complete attack path of the attacker and identify the key cyber and physical components that are vulnerable in the electric power CPS network. This method proposes the concept of coordinated network attack modes and uses corresponding algorithms for physical attack event identification, cyber attack sequence identification, and multi-step frequent attack mode extraction.

This paper comprehensively considered the attack process of both cyber and physical space. Moreover, the proposed method does not rely on various complicated association rules that are set manually, and does not require a large amount of prior knowledge to achieve good practical results. In complex electric power CPS networks, the attack patterns discovered in different network cooperative attack sequences may be local. Therefore, future research will focus on the investigation of attack pattern fusion methods in large-scale topologies.

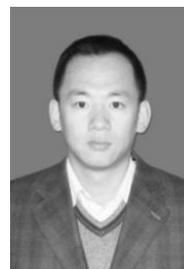

**LEI WANG** was born in Jilin, China. He is pursing his Doctor's degree of electric engineering in Northeast Electric Power University . He is an Associate professor at the School of Information Engineering. His research interest is Cyber-physical system attack and identification in Smart grid. He won the first prize of scientific and technological progress in Jilin Province in 2019 .

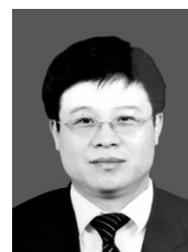

**ZHAOYANG QU** received his Ph.D degree in electric engineering from China Northeast Electric Power University in 2006 and 2010,and his M.S degree in Dalian University of Technology in 1984 and 1988. He is currently a professor and doctoral tutor in the School of Information Engineering of Northeast Electric Power University. His interests include smart grid and power information processing, virtual reality and network technology.

He is a member of the China Electric Engineering Society Power Information Committee, the vice president of the Jilin Province Image and Graphics Society, the head of the Power Big Data Intelligent Processing Engineering Technology Research Center, and the Jilin Governor Baishan Scholar. He is also a top-notch innovative talent in Jilin Province and a young and middle-aged professional and technical talent with outstanding contributions. He presided over the completion of 2 national natural science funds, won the second prize of Jilin Province Science and Technology Progress Award, and wrote 46 papers on electric power information SCI/EI search.

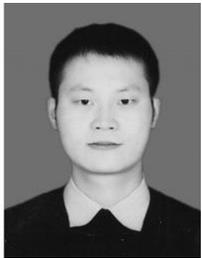

**YANG LI** (S'13–M'14–SM'18) was born in Nan yang, China. He received his Ph.D. degree in Electric Engineering from North China Electric Power University (NCEPU), Beijing, China, in 2014. He is an Associate professor at the School of Electric Engineering, Northeast Electric Power University, Jilin, China. Currently, he is also a China Scholarship Council (CSC)-funded post doc with Argonne National Laboratory, Lemont, United States. His research interests include power system stability and control, integrated energy system, renewable energy integration, and smart grids. Dr. Li is the Associate Editor of IEEE Access, IET Renewable Power Generation, and Electrical Engineering.

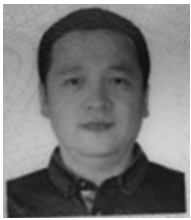

**KEWEI HU** was born in May 1968, he is now working in the Power Dispatching Control Center of the State Grid Jilin Electric Power Co., Ltd., and graduated from Northeast Electric Power University with a master's degree in power system and automation. His research direction is power system dispatching automation.

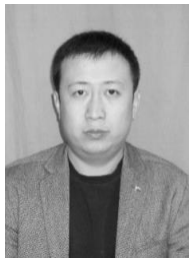

**JIAN SUN** received the B.S. degree from the Northeast Electric Power University, in 2001. He is currently a Senior Engineer at State Grid Jilin Power Co., Ltd. Baishan Power Supply Company. His research interest is safety monitoring in the smart grid.

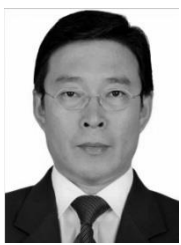

**KAI XUE** received the M.S. degree from the Northeast Normal University. He is currently a Senior Engineer at State Grid Jilin Power Co., Ltd. Training Center. His research interest is power system automation.

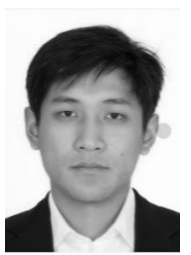

**MINGSHI CUI** received the B.S. degree from the North China Electric Power University, in 2010.
He is currently an Engineer at State Grid Inner Mongolia Eastern Electric Power Company. His research interest is information processing in smart grid.